\documentclass[reprint,nofootinbib,onecolumn,superscriptaddress,a4paper]{revtex4-2}
\usepackage{color}
\definecolor{darkgreen}{rgb}{0,0.35,0}

\usepackage{hyperref}
\usepackage{amsfonts}
\usepackage{amsmath}
\usepackage{bbold}
\usepackage[noabbrev]{cleveref}
\usepackage{subfig}
\usepackage{mathtools,slashed,tensor}
\usepackage[font=scriptsize,labelfont=bf]{caption}
\usepackage{float}
\usepackage[top=2cm, bottom=2cm, left=3cm, right=3cm]{geometry}
\usepackage{tikz}
\usepackage{siunitx}
\usetikzlibrary{decorations.pathmorphing, arrows, patterns}
\usetikzlibrary{positioning}
\newcommand{\be}{\begin{equation}}

\newcommand{\ee}{\end{equation}}
\newcommand{\bea}{\begin{eqnarray}}
\newcommand{\eea}{\end{eqnarray}}
\newcommand{\bef}{\begin{figure}}
\newcommand{\eef}{\end{figure}}
\newcommand{\bce}{\begin{center}}
\newcommand{\ece}{\end{center}}
\newcommand{\ii}{\ensuremath{\mathrm{i}}}
\def\lsim{\mathrel{\rlap{\lower4pt\hbox{\hskip1pt$\sim$}}
    \raise1pt\hbox{$<$}}}         %less than or approx. symbol
\def\gsim{\mathrel{\rlap{\lower4pt\hbox{\hskip1pt$\sim$}}
    \raise1pt\hbox{$>$}}}         %greater than or approx. symbol
\usepackage{tikz}
\usetikzlibrary{decorations.pathmorphing, arrows, patterns}
\tikzset{snake it/.style={-stealth,
decoration={snake,
    amplitude = .4mm,
    segment length = 2mm,
    post length=0.9mm},decorate}}

\newcommand{\ucharles}{Faculty of Mathematics and Physics, Charles University, V Hole\v{s}ovi\v{c}k\'ach 2, 18000 Prague 8, Czech Republic}
\newcommand{\uach}{Instituto de Ciencias F\'isicas y Matem\'aticas, Universidad Austral de Chile, Casilla 567, 5090000 Valdivia, Chile}

\newcommand{\borisaff}{Wöhlergasse 6, 1100 Vienna, Austria}
\begin{document}

\title{Turning graphene into a lab for noncommutativity}
\author{Alfredo Iorio}
\email{iorio@matfyz.cuni.cz}
\affiliation{\ucharles}
\author{Boris Iveti\'{c}}
\email{bivetic@yahoo.com}
\affiliation{\borisaff}
%\author{S.~Mignemi}
%\email{smignemi@unica.it}
%\affiliation{\infn}
\author{Pablo Pais}
\email{pais@matfyz.cuni.cz}
\affiliation{\uach}
\affiliation{\ucharles}

\begin{abstract}
It was recently shown that taking into account the granular structure of graphene lattice, the Dirac-like dynamics of its quasiparticles resists beyond the lowest energy approximation. This can be described in terms of new phase-space variables, $(\vec{X},\vec{P})$, that enjoy generalized Heisenberg algebras. In this letter, we add to that picture the important case of noncommuting $\vec{X}$, for which $[X^i,X^j] = \ii \, \theta^{i j}$ and we find that $\theta^{i j} = \ell^2 \,  \epsilon^{i j}$, with $\ell$ the lattice spacing. We close by giving both the general recipe and a possible specific kinematic setup for the practical implementation of this approach to test noncommutative theories in tabletop analog experiments on graphene.
\end{abstract}

\maketitle

%%%%%%%%%%%%%%%%%%%%%%%%%%%%%%%%%%%%%%%%%%%%%%%%%%%%%%%%%%%%%%%%%%%%%%%%%%%%%%%%%%%%%%%%%%%%%%%%%%%%%%%%%%%%%%%%%%%%%%%%%%
\section{Introduction} \label{intro}

In this letter, in the same spirit of \cite{threelayers}, that is, regarding graphene as an analog of certain quantum gravity (QG) scenarios, we consider the possibility that coordinates enjoy canonical noncommutativity (CNC)
\begin{equation}\label{canonical}
  [X_i, X_j] = \ii \, \theta_{i j},
\end{equation}
with $\theta_{ij}$ a constant antisymmetric tensor of dimension $L^2$ and $i,j, ... = 1,2$.

With analogs we can test the behaviour of one physical system, by doing experiments on another physical system \cite{Feynman}. This becomes particularly important when we want to test theories describing phenomena that otherwise lie far beyond our experimental capabilities \cite{Xons}.

The research on gravity analogs can be seen as originating from the seminal work of Unruh~\cite{UnruhAnalog}. He proposed searching for experimental signatures of the Unruh \cite{bill} and the Hawking effect \cite{haw0}, in a fluid-dynamical analog. These days, we can reproduce many different aspects of fundamental physics in analog systems, in Bose--Einstein condensates~\cite{MunozdeNova2019}, in Weyl semimetals~\cite{Gooth2017}, in graphene \cite{i2,weylgraphene,iorio2012,iorio2014,iorio2015}, and on many more platforms \cite{Barcelo2005}.

Our interest here is on graphene as an analog of high-energy fundamental physics, based on the fact that its low-energy quasiparticle excitations~\cite{PacoReview2009} are massless Dirac pseudo-relativistic fermions (the matter fields $\psi$), propagating in a carbon two-dimensional honeycomb lattice. We shall focus on a regime that goes slightly beyond the lowest energy regime just recalled, where nonlinear effects start playing a role, and the modified structure of space-time of QG may be mimicked.

As it was first recognised in \cite{egypt2018,ip5}, the effective high-energy momenta in combination with the standard (low-energy) spatial coordinates obey a modified Heisenberg algebra, characterised by deformed commutation relations between position and momenta, which is characteristic of certain QG models involving a minimal length (MLM) \cite{AliDasVagenas1,*AliDasVagenas2,*AliDasVagenas3,*AliDasVagenas4,*AliDasVagenas5}.

A recent paper  \cite{threelayers} continued on this, allowing the spatial coordinates at high energy to be modified as well. One crucial result there is that, in two important cases with GUPs, such high-energy coordinates simply coincide with the low-energy coordinates. In other cases, such coordinates have complicated expressions regarding the standard, measurable phase space variables. In all cases, though, coordinates commute. Here, we show that noncommuting coordinates are indeed possible in a canonical form, putting on more firm grounds what was first speculated in \cite{ip5}.

CNC is one of the earliest models of noncommutativity, appearing independently in various high-energy physics theories \cite{Doplicher1995,SeibergWittenNC,Madore2000,iorio2002noncommNoether,*iorio2008noncommcomment}. For instance, in \cite{Doplicher1995}, it is argued that the uncertainty relations descending from (\ref{canonical}) may reflect constraints on the localization of a particle due to classical gravity when the distance becomes less then the Schwarzschild radius. The idea there is that there is a value of the wavelength of the photon, $\lambda_\gamma$, used to measure the position of a particle, that coincides with the Schwarzschild radius, $R_\gamma$, associated with that energy, $E_\gamma = \hbar \omega_\gamma$. When such a length scale is reached, one cannot go beyond it. Hence, there is a minimal length, which gives a minimal spatial uncertainty. In \cite{Doplicher1995}, and usually, such length is Planck's, but there are also alternatives \cite{iorio2020generalized}. In this letter, as we shall see in a moment, we do not need such arguments as the minimal length is there regardless of any gravitational radius (on this, see also \cite{iorio2020generalized}).

In CNC models, the algebra is usually given by (\ref{canonical}) and the two standard commutation relations, between momenta and momenta and between positions and momenta. In the case of graphene, however, in addition to  (\ref{canonical}), the commutation relation between positions and momenta changes as well to
\begin{equation}\label{canonicalXP}
  [X_i, P_j] = \ii \, \hbar F_{i j}(P),
\end{equation}
where $F_{i j}(P)$ is an antisymmetric tensor that shares many features with those appearing in MLMs. This leads to a novel type of noncommutativity, which is a mix of CNC and MLM.

The letter is organized as follows. Section \ref{GrapheneQG} is dedicated to recalling the main reasons why graphene is a QG analog and setting the notation. Section \ref{Sect_from_second_to_third_layer} is where we obtain noncommuting coordinates. Section \ref{recipe} is where we explain, in general terms, how to use graphene as a tabletop laboratory to test noncommutative theories, while in Section \ref{Gedanken}, we focus on a specific kinematic arrangement. Section \ref{Conclusions} is dedicated to our conclusions.

%%%%%%%%%%%%%%%%%%%%%%%%%%%%%%%%%%%%%%%%%%%%%%%%%%%%%%%%%%%%%%%%%%%%%%%%%%%%%%%%%%%%%%%%%%%%%%%%%%%%%%%%%%%%%%%%%%%%%%%%%%
\section{The set-up}
\label{GrapheneQG}
%%%%%%%%%%%%%%%%%%%%%%%%%%%%%%%%%%%%%%%%%%%%%%%%%%%%%%%%%%%%%%%%%%%%%%%%%%%%%%%%%%%%%%%%%%%%%%%%%%%%%%%%%%%%%%%%%%%%%%%%%%

The dispersion relations of graphene can be approximated by \cite{threelayers}
\bea
  E_\pm = \eta \left( \pm |{\cal F}| - 0.15 |{\cal F}|^2 \right)  \label{DispRelModified}
\eea
where $\eta \simeq - \SI{2.8}{\electronvolt}$ is the nearest neighbour hopping energy,
\begin{equation}\label{firstfunction}
  {\cal F} ({\vec{k}}) = \sum_{m = 1}^{3} e^{(\ii/\hbar) {\vec{k}} \cdot {\vec{s}^{(1)}}_m} =
  e^{- \ii \ell k_2/\hbar} \left[ 1 + 2 e^{\ii \frac{3}{2} \ell k_2/\hbar} \cos\left(\frac{\sqrt{3}}{2\hbar} \ell k_1\right) \right] \,,
\end{equation}
$\ell$ is the lattice spacing, the three vectors ${\vec{s}^{(1)}}_m$ connect nearest neighbouring lattice sites \cite{egypt2018} (all vectors are spatial two-dimensional), and the zero of the energy is set at the Dirac point, i.e. $E_\pm|_{k_D} = 0$.

Expanding around the Dirac points, see \cite{PacoReview2009}, at the leading order in the parameter $\ell |\vec p|/\hbar$ (the long-wavelength limit), one gets
\begin{equation}\label{lin dis rel}
   E_\pm = \pm v_F  |\vec{p}| \,,
\end{equation}
where $\vec{p}$ is a small momentum and $v_F \equiv 3/2 \, \eta \ell / \hbar \simeq c / 300$, is the Fermi velocity. Therefore, the physics of the conductivity quasiparticles of graphene is governed, at order $O(|\vec{p}|)$, by the standard Dirac linear Hamiltonian
\begin{equation}
H (p) = v_{F}\,\sum_{p}\,\psi^{\dagger}_{p} \, \slashed{p} \, \psi_{p} \,,\label{Dirac_Hamiltonan}
\end{equation}
that, through
\begin{equation}\label{SecEq}
{\rm det} (H - E\mathbb{1}) = 0 \,,
\end{equation}
gives (\ref{lin dis rel}) as dispersion relation.  Our convention is\footnote{There are many conventions related to different choices for the pairs of inequivalent Dirac points and various arrangements of the $a$ and $b$ operators to form the spinor $\psi$. See Appendix B of \cite{ip3} for more details.} $\psi_{p}^{\dagger}=(b_{p}^{*}, a_{p}^{*})$, where $a_{p}$ and $b_{p}$ are the annihilation operators for the $L_A$ and $L_B$ sublattices \cite{PacoReview2009}.

Once one gets away from the Dirac points, higher-order momenta enter, and the dispersion relation is no longer linear. It has been recognised in \cite{egypt2018} that one can maintain a linear dispersion relation, provided one redefines the momenta. The relation (\ref{DispRelModified}) defines the auxiliary quantity
\be\label{supermomenta}
{\vec{P}_0} \equiv - \frac{\hbar}{\ell} \, \left( {\rm Re} {\cal F}, {\rm Im} {\cal F} \right) \,,
\ee
that was dubbed \textit{supermomentum} in \cite{threelayers}.

The dispersion relation (\ref{DispRelModified}) can be obtained from the secular equation (\ref{SecEq}), written for the following Hamiltonian given in terms of the supermomenta $\vec{P_{0}}$
\begin{equation}\label{newDiracHwithP0}
  H (P_0) = V_F  \sum_{\vec{k}} \psi^\dag_{\vec{k}} \left(\not\!P_0 - A \; \not\!P_0 \not\!P_0 \right) \psi_{\vec{k}} \,,
\end{equation}
where $V_F = \eta \ell / \hbar = 2/3 \, v_F$, $\slashed{P}_0 \equiv \vec{\sigma} \cdot {\vec{P}_0}$, $\vec{\sigma}$ are the usual Pauli matrices and $A \simeq 0.15 \ell / \hbar$.

Noticeably, this is the same Hamiltonian obtained in the phenomenology of QG, when generalizing the Dirac Hamiltonian to accommodate a GUP with a minimal fundamental length \cite{AliDasVagenas1,*AliDasVagenas2,*AliDasVagenas3,*AliDasVagenas4,*AliDasVagenas5}. Here such a fundamental length is clearly given by the carbon-to-carbon distance, $\ell$.

In \cite{AliDasVagenas1,*AliDasVagenas2,*AliDasVagenas3,*AliDasVagenas4,*AliDasVagenas5}, the authors introduce \textit{two} different ``layers'' describing the given physical system. Namely, ``high energy'' and ``low energy'' phase-space variables that, adjusting our notation to theirs, we may call $(\vec{X},\vec{P})$ and $(\vec{X}_0, \vec{P}_0)$, respectively. Here (and there), the definition of the ``high-energy momenta'' descends directly from (\ref{newDiracHwithP0}), that in turn is associated with the dispersion relations (\ref{DispRelModified})
\begin{equation}\label{hypermomenta}
{\vec{P}} \equiv {\vec{P}_0} (1 - A |\vec{P_0}| ) \,.
\end{equation}
This we call the \emph{hypermomentum}.

With this definition, we arrive again at the standard Dirac linear Hamiltonian
\begin{equation}
H (P) = V_{F}\,\sum_{k}\,\psi^{\dagger}_{k} \, \slashed{P} \, \psi_{k} \;,\label{Hamiltonan_hypermomenta}
\end{equation}
with dispersion relation
\begin{equation}\label{lin dis rel_P}
   E_\pm = \pm V_F  |\vec{P}| \;.
\end{equation}

Notice that, to obtain $H(P_0)$ from $H(P)$, one should first use the expansion (\ref{hypermomenta}), and then use the customary Dirac prescription $|\vec{P_{0}}| \to \slashed{P_{0}}$
\begin{eqnarray}
H (P(P_0)) & = & V_{F}\,\sum_{k}\,\psi^{\dagger}_{k} \, \vec{\sigma} \cdot \vec{P} \, \psi_{k} \nonumber \\
           & = & V_{F}\,\sum_{k}\,\psi^{\dagger}_{k} \vec{\sigma} \cdot \left(\vec{P_{0}}(1 - A |\vec{P_{0}}|)\right) \, \psi_{k} \nonumber \\
           &\to& V_{F}\,\sum_{k}\,\psi^{\dagger}_{k} \left( \vec{\sigma} \cdot \vec{P_{0}} - A (\vec{\sigma} \cdot \vec{P_{0}}) \, (\vec{\sigma} \cdot \vec{P_{0}}) \right) \, \psi_{k} \nonumber \\
           & = & V_{F}\,\sum_{k}\,\psi^{\dagger}_{k} \left(\slashed{P_{0}} - A |\vec{P_{0}}|^2 \right) \, \psi_{k} = H (P_0) \,,\label{Hamiltonan_hyper_supermomenta}
\end{eqnarray}
where, in the second term of the last line, we used $\sigma_i \sigma_j = \delta_{i j} \mathbb{1} + i \epsilon_{i j k} \sigma_k$ and the symmetry of $P^i_0 P^j_0$.

For all the above, see \cite{threelayers}. There it is shown that, in this set-up, we do know supermomenta and hypermomenta in terms of the measurable momenta $\vec{p}$
\begin{equation}\label{momenta}
P_{0}^{i}(p) \,\,\,\, \mbox{and} \,\,\,\, P^i (P_{0}(p)) \,,
\end{equation}
and one needs to find supercoordinates and hypercoordinates in terms of measurable phase-space variables $(\vec{x},\vec{p})$
\begin{equation}\label{coordinates}
X_{0}^{i}(x,p) \,\,\,\, \mbox{and} \,\,\,\, X^i (X_0 (x,p), P_{0}(p)) \,,
\end{equation}
where, for the latter, we can take as an experimental fact that the phase-space variables are canonical
\begin{equation}\label{commutators_x_p}
[x^{i},p^{j}]= \ii\hbar\delta^{ij}  , \ \ \ [x^{i},x^{j}] = 0 = [p^{i},p^{j}] \;.
\end{equation}

Henceforth, we shall always have
\begin{equation}\label{PP}
  [P_0^i (p),P_0^j(p)] = 0 = [P^i(p),P^j(p)] \,,
\end{equation}
whereas a great variety of choices are allowed for the associated generalized coordinates.

%%%%%%%%%%%%%%%%%%%%%%%%%%%%%%%%%%%%%%%%%%%%%%%%%%%%%%%%%%%%%%%%%%%%%%%%%%%%%%%%%%%%%%%%%%%%%%%%%%%%%%%%%%%%%%%%%%%%%%%%%%
\section{Noncommuting Hypercoordinates}
\label{Sect_from_second_to_third_layer}
%%%%%%%%%%%%%%%%%%%%%%%%%%%%%%%%%%%%%%%%%%%%%%%%%%%%%%%%%%%%%%%%%%%%%%%%%%%%%%%%%%%%%%%%%%%%%%%%%%%%%%%%%%%%%%%%%%%%%%%%%%

Upon introduction of the hypermomenta $P^i$ in \cref{hypermomenta}, $P^{i}=P_{0}^{i}(1-A \, |\vec{P_{0}}|)$, which, in turn, can be expressed in terms of standard $p^{i}$ as \cite{threelayers}
\begin{align}\label{HypermomentaCartesian}
P^{1} & = p^{1} (1-A\,|\vec{p}|) + \ell \left(\frac{A\, (p^{1})^2 \left((p^{1})^{2} -3 (p^{2})^2\right)}{4 p}+\frac{1}{4} (1-A \,|\vec{p}|) \left((p^{2})^2-(p^{1})^2\right)\right)-\frac{1}{8} \,\ell^2 \,|\vec{p}|^2\, p^{1} \;, \\
P^{2} &= p^{2} (1-A \,|\vec{p}|) + \frac{\ell\, p^{2} \left(-A (p^{1})^3 - 5 A p^{1} (p^{2})^2+2 |\vec{p}| p^{1}\right)}{4 |\vec{p}|}-\frac{1}{8} \,\ell^2 \,|\vec{p}|^2 p^{2} \;, \nonumber
\end{align}
the question is to find  hypercoordinates $X^i (x,p)$ that complement $P^i$ in a way that closes the Heisenberg algebra. As discussed at length in \cite{threelayers}, unlike the momenta $P^i(p)$, whose functional dependence on the ``true'' (laboratory) momenta $p^i$ is dictated by the dispersion relation, the functional form of the hypercoordinates on the laboratory phase-space variables is not uniquely fixed by the algebraic requirements alone. Two possible definitions were explored in \cite{threelayers}. Here we consider another, more general, case:
\begin{equation}\label{hyper_algebra}
[X^{i},P^{j}] =  \ii \hbar\, F^{i\,j}(\vec{P}) \;, \ \  [X^{i},X^{j}] =\ii G^{i\,j} (\vec{P})  \;, \ \   [P^{i},P^{j}] = 0 \;,
\end{equation}
where
\begin{equation} \label{CallFij}
F^{i\,j}(\vec{P}) \equiv f^{i\,k}(\vec{P})\,\left((1-A\,|\vec{P}|-A^{2}\,|\vec{P}|^{2})\delta^{kj}-A\,\frac{P^{k}\,P^{j}}{|\vec{P}|}\,(1+A\,|\vec{P}|)\right) \;,
\end{equation}
with
\begin{equation}\label{Fij(P0)}
  f^{i j} (\vec{P}) = \delta^{i j}
  + \frac{1}{2} \, \ell \, (1 + A|\vec{P}|) \left(
                  \begin{array}{cc}
                  - P^1 & P^2 \\
                    P^2 & P^1 \\
                  \end{array}
                \right)
  - \frac{1}{2} \, \ell^2 \left(
                  \begin{array}{cc}
                    (P^1)^2  & P^1 P^2 \\
                    P^1 P^2 &  (P^2)^2   \\
                  \end{array}
                \right) \;,
\end{equation}
being $G^{i j}$ an arbitrary antisymmetric function of $\vec{P}$. Besides zero, the simplest choice for $G^{i j}$ is
\begin{equation}\label{Hij}
  G^{i j} = L^2 \, \epsilon^{ij} \,,
\end{equation}
where $\epsilon^{ij}$ is the totally antisymmetric tensor in two dimensions and $L$ is a length scale. Since these calculations are $O(\ell^2)$, length scales of higher orders could not be appreciated, henceforth
\begin{equation}\label{L(l)}
  L (\ell) = a \, \ell \;,
\end{equation}
where $a$ is a $O(1)$ numerical constant that we set to $1$, for simplicity.

Note that this is the only possibility compatible with our requirements, apart from the choice of \cite{threelayers}, $G^{i j}=0$. If one wants a result proportional to $|\vec{P}|$, this requires a factor of $\ell^3$ to compensate for the physical dimension, which is bigger than $O(\ell^{2})$. If we include $\ell/|\vec{P}|$, we do have the correct physical dimension, but $G^{ij}$ would not have a reasonable limit for $|\vec{P}| \to 0$.

Therefore we have
\begin{equation}\label{X0X0H}
  [X^i,X^j] = \ii \, \theta^{ij} \,,
\end{equation}
where
\begin{equation}
\theta^{i j} = \, \ell^2 \, \epsilon^{ij} \,.
\end{equation}
This is the CNC we were looking for, realized in terms of the commutative phase-space variables, $(\vec{x},\vec{p})$, by
$X_{1} = x_{1} - 1 / (2 \hbar) \, \ell^2 \, p_{2}$ and $X_{2} = x_{2} + 1 / (2 \hbar) \, \ell^2 \, p_{1}$, i.e.,
\begin{equation} \label{XNC(x,p)}
X^{i} = x^{i} - \frac{1}{2 \hbar} \, \theta_{i j} p^{j} \,.
\end{equation}
This realization is known in the literature as the Bopp's shift, see, e.g., \cite{Curtright1997}.

%%%%%%%%%%%%%%%%%%%%%%%%%%%%%%%%%%%%%%%%%%%%%%%%%%%%%%%%%%%%%%%%%%%%%%%%%%%%%%%%%%%%%%%%%%%%%%%%%%%%%%%%%%%%%%%%%%%%%%%%%%
\section{Operational recipe for the analog noncommutativity}
\label{recipe}
%%%%%%%%%%%%%%%%%%%%%%%%%%%%%%%%%%%%%%%%%%%%%%%%%%%%%%%%%%%%%%%%%%%%%%%%%%%%%%%%%%%%%%%%%%%%%%%%%%%%%%%%%%%%%%%%%%%%%%%%%%

We now want to give the operational recipe to use graphene to test noncommutative theories in an analog lab. Notice that this scheme could also be applied to other analog systems.

First, we have to bear in mind that we only have experimental access to the measurable ``lab'' variable $(\vec{x},\vec{p})$. The system is described by a Hamiltonian $H(p)$, written in terms of $(\vec{x},\vec{p})$. This Hamiltonian is very complicated, as it must include higher orders; see \cite{threelayers,egypt2018} and \eqref{Hamiltonian_total_Op3} later here. However, this is not an issue because we have access to experiments. Hence, ``nature makes the computations'', so it should be possible to compute everything, at least numerically. To this, $H(p)$ corresponds to $H(P)$ in (\ref{Hamiltonan_hypermomenta}), where the variables are the noncommutative $\vec{X}$.

With these, the operational recipe is
\begin{itemize}
  \item Measure observables in the lab, such as the spectrum, $E_n (x,p)$. We might do so for the free system as a test and then apply a suitably engineered potential, $V(x,p)$. Such potential should correspond, through $x_i(X,P)$ and $p_i (P)$ (see \eqref{XNC(x,p)} and \eqref{HypermomentaCartesian}), to a wanted potential, $V(X)$, in the ``target noncommutative system''.
  \item Rewrite the outcomes of that measurement in the noncommuting variables, e.g.,\\
  $E_n [x(X,P), p (X,P)]$, for the example of the spectrum.
  \item Perform a calculation using the Hamiltonian $H(P)$, first for the free case and then adding the potential $V(X)$ to it. The free case is only for setting up/verifying the correspondence. We want to study the interacting case to test the model based on noncommuting $\vec{X}$. We then obtain the observables, e.g., $E_n(X,P)$. Here, we are entirely theoretical/mathematical.
  \item Compare the measured observables with the theoretical prediction based on the noncommutative theory. E.g., $E_n [x(X,P), p (X,P)]$, obtained experimentally, with $E_n(X,P)$, obtained from the noncommutative theory.
\end{itemize}

Notice that we shall have two sets of data written in the $(\vec{X},\vec{P})$ variables: one set is obtained by rewriting the results of the experiments and calculations done with the setting $H(p)$, $(\vec{x},\vec{p})$. The other set is obtained by performing the (noncommutative) calculations within the setting $H(P)$, $(\vec{X},\vec{P})$.

The above recipe can be sketched in the following diagram

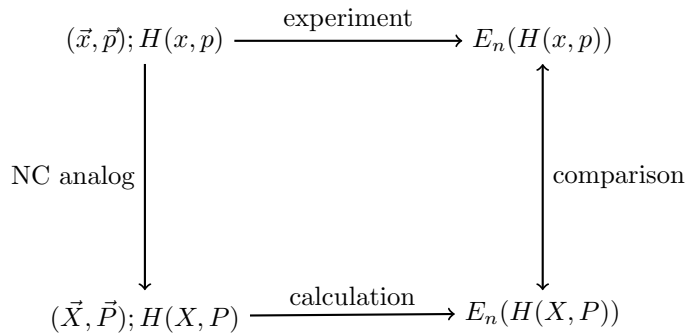
\begin{center}
    \begin{tikzpicture}[node distance = 3cm, thick]
        \node (1) {$(\vec{x},\vec{p}); H(x,p)$};
        \node (2) [right=of 1] {$E_n(H(x,p))$};
        \node (3) [below=of 1] {$(\vec{X},\vec{P}); H(X,P)$};
        \node (4) [below=of 2] {$E_n(H(X,P))$};
        \draw[->] (1) -- node [midway,above] {experiment} (2);
        \draw[->] (1) -- node [midway,left]{NC analog} (3);
        \draw[<->] (2) -- node [midway,right] {comparison} (4);
	\draw[->] (3) -- node [midway,above] {calculation} (4);
    \end{tikzpicture}
\end{center}

For instance, for the free case (no external potential), a typical Hamiltonian that departs from the linear approximation, given in lab variables $(\vec{x},\vec{p})$, is \cite{threelayers}
\begin{eqnarray}\label{Hamiltonian_total_Op3}
H & = & v_F \sum_{\vec{p}} \psi^\dag_{\vec{p}} \Bigg{[} \sigma_1 \left( p_1 - \frac{\ell}{4} (p^2_1 - p^2_2) - \frac{\ell^2}{8} p_1 (p^2_1 + p^2_2) \right)  +  \sigma_2 \left( p_2 + \frac{\ell}{2} p_1 p_2 - \frac{\ell^2}{8} p_2 (p^2_1 + p^2_2) \right) \nonumber \\
  & - & \frac{3}{2} A \left( (p^2_1 + p^2_2) - \frac{\ell}{2} p_1^3 + \frac{3 \ell}{2} p_1 p^2_2 \right) \Bigg{]}  \psi_{\vec{p}} \;.
\end{eqnarray}
To this corresponds the free Hamiltonian (\ref{Hamiltonan_hypermomenta}), $H (P) = V_{F}\,\sum_{k}\,\psi^{\dagger}_{k} \, \slashed{P} \, \psi_{k}$, written in terms of the hypervariables.

%%%%%%%%%%%%%%%%%%%%%%%%%%%%%%%%%%%%%%%%%%%%%%%%%%%%
\section{Applying the recipe in a specific setup} \label{Gedanken}
%%%%%%%%%%%%%%%%%%%%%%%%%%%%%%%%%%%%%%%%%%%%%%%%%%%%

The construction of physical theories pertaining to the deformed commutator in (\ref{X0X0H}) has a long and rich tradition, see, e.g., \cite{Madore2000,Jackiw,SeibergWittenNC,IorioDICE2006},  among many other references. Such noncommutativity relies on the mapping between algebras of functions of the commuting coordinates (standard) and functions of noncommuting coordinates. The quintessential ingredient is the commutator (\ref{X0X0H}) itself \cite{Madore2000}.

In this section, we shall provide an example of a possible application of the recipe. Given the generality of the results of this letter, we shall not do that by focusing on a full-fledged quantum electrodynamic calculation. The latter would need to face the subtleties of having to deal with massless particles, an issue that points to the Klein paradox, somehow faced in graphene with deformed algebras already, albeit in a commutative context \cite{Naveed2022}. Future work on the noncommutative Klein paradox, within the approach presented here, has been planned \cite{WIP}.

What we shall do, instead, is to focus on kinematics, which is the most straightforward way to test the novel noncommutativity introduced in this work, Eq. (\ref{X0X0H}). This does not require the application of the full dynamics of the deformed field theory to describe the processes. We only need to recognize that position operators represent generators of infinitesimal displacements on the momentum space. Since they do not commute, this also implies that finite displacements of the momenta will only commute when collinear, but in general, they do not commute. The action of the displacement of electrons' momentum represents the absorption or emission of a photon. With this in mind, the example we single out here will convey general information common to many dynamical theories.

The noncommutativity then can be observed experimentally in the following way. Let us assume an electron with a fixed initial value of the momentum, $\vec{P}$, absorbs two non-collinear photons  momenta $\vec{Q}$ and $\vec{Q'}$, in succession. In the commutative case, the final value of electron momentum, $\vec{K}$, does not depend on the order in which the two photons are absorbed. In the noncommutative case, the ordering will make a difference due to the noncommutativity of the momenta addition law, which is a direct consequence of (\ref{X0X0H}). Thus, the two final momenta of the electron $\vec{K}$ and $\vec{K'}$ are generally different for the processes depicted in the figure below.

\begin{figure}[h]
\begin{center}
\begin{tikzpicture}[thick]
\draw[->] (-5,0) -- (-2,0);
  \draw [snake it]
    (-5,1) -- (-3.5,0);
  \draw [snake it]
    (-4,-1) -- (-2.5,0);
\draw (-5.7,0) node[anchor= west] {$\vec{P}$};
\draw (-5.7,1.1) node[anchor= west] {$\vec{Q}$};
\draw (-4.7,-1.1) node[anchor= west] {$\vec{Q'}$};
\draw (-2,0) node[anchor= west] {$\vec{K}$};
\draw[->] (2,0) -- (5,0);
  \draw [snake it]
    (2,1) -- (3.5,0);
\draw [snake it]
    (3,-1) -- (4.5,0);
\draw (1.3,0) node[anchor= west] {$\vec{P}$};
\draw (1.3,1.1) node[anchor= west] {$\vec{Q'}$};
\draw (2.3,-1.1) node[anchor= west] {$\vec{Q}$};
\draw (5,0) node[anchor= west] {$\vec{K'}$};
\end{tikzpicture}
\caption{An electron of the initial momentum $\vec{P}$ absorbs two photons of the momenta $\vec{Q}$ and $\vec{Q'}$ in the opposite order in diagrams on the left and the right. The resulting momenta of the electron are, in general, different in the two cases, $\vec{K}\neq\vec{K'}$.}
\label{process}
\end{center}
\end{figure}
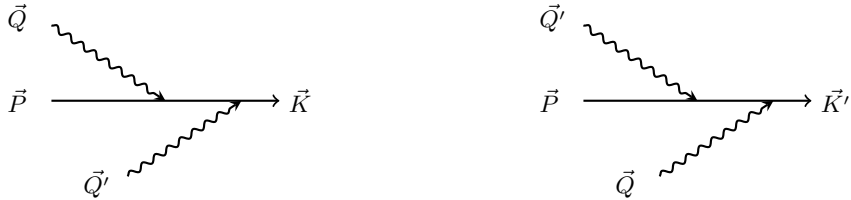

The experimental setting required to achieve this is the following. One needs two sets of electrons prepared in the same initial state of the definite momenta $\vec{P}$, which are exposed to two consecutive short monochromatic non-collinear electromagnetic pulses. The interaction of light with the $\pi$-electrons of graphene is well-studied these days, a noticeable example being the theoretical and experimental studies on the laser-graphene interaction \cite{Higuchi2017-dp}, see also \cite{tloopPRD2020}.

The order of the shooting of photons on the two sets must be as in Fig. \ref{process}. Finally, one measures the difference between the momenta of the electrons in the two final sets.

Now, the general form of the momentum addition law is not familiar for the noncommutativity defined by relation (\ref{X0X0H}). One can, however, on purely dimensional grounds, determine the leading order form
\begin{equation}
\vec{Q} \oplus \vec{P} = (1+a_1\theta_{jk}P_j Q_k) \vec{P} + (1+a_2\theta_{jk}P_j Q_k) \vec{Q} \,.
\end{equation}
The symbol $\oplus$ represents the fact that the momentum conservation law gets deformed; we follow a convention where the momentum on the left displaces the momentum on the right; that is, the electron of the momentum $\vec{P}$ absorbs (gets displaced by) the photon of the momentum $\vec{Q}$.\footnote{The development of the field theory of identical particles, such as scalar $\phi^n$ theory, on the noncommutative background, suffers from insurmountable ambiguities related to their indistinguishability. This, however, does not represent a physical constraint since no known elementary particles emit or absorb like particles.} The difference in the coefficients $a_1$ and $a_2$ is the measure of noncommutativity of the momenta addition law. For the two processes described above, we have then,
\begin{align*}
\vec{K} & = \vec{Q'} \oplus (\vec{Q} \oplus \vec{P}) \;,
\end{align*}
and
\begin{align*}
\vec{K}' &= \vec{Q} \oplus (\vec{Q'} \oplus \vec{P} )\;.
\end{align*}
After some algebra, we can express the difference $\vec{K'}-\vec{K}$ in a convenient expression that, in vectorial notation, reads
\begin{equation}\label{K-difference}
\vec{K'} - \vec{K} = \ell^{2} \, (Q'_{1}\,Q_{2} - Q_{1}\,Q'_{2})\,\left(3a_{1}\vec{P} + (a_{1}+a_{2})(\vec{Q} + \vec{Q'})\right) \;.
\end{equation}
It is clear from Eq. (\ref{K-difference}) that the difference is zero when $\vec{Q}$ and $\vec{Q'}$ are collinear and that the leading order correction in the length parameter is $\ell^{2}$.

Now, the crucial part of the story: Suppose that the light interacting with graphene can probe nonlinear regimes of the dispersion relations for the $\pi$-electrons, as discussed in \cite{tloopPRD2020}. This nonlinearity, within the framework given here, see also \cite{threelayers}, can be transferred to the new phase space variables, $(\vec{X}, \vec{P})$, with $\vec{X}$ given by (\ref{XNC(x,p)}), and noncommuting, as opposed to the phase space variables, $(\vec{x}, \vec{p})$, which we actually measure in the lab. Suppose the results of the above-described experiments are now written in terms of $(\vec{X}, \vec{P})$. In that case, we can check whether the predictions of the theoretical arguments given above, based on noncommutativity, indeed occur!

It is vital to notice two facts. First, if we use $(\vec{x}, \vec{p})$, we shall only see some nonlinear effects, and no noncommutativity can be there. Second, even in the new variables,  $(\vec{X}, \vec{P})$, it is by no means trivial that we see the noncommutativity. We may as well not see any noncommutativity at all, even in the new variables! It is then a real test of noncommutativity that we are performing by a clever change of phase space coordinates, in part dictated by the detailed form of the dispersion relations of the material, in part dictated by the presence of a length scale, given by the lattice spacing, $\ell$.

%%%%%%%%%%%%%%%%%%%%%%%%%%%%%%%%%%%%%%%%%%%%%%%%%%%%%%%%%%%%%%%%%%%%%%%%%%%%%%%%%%%%%%%%%%%%%%%%%%%%%%%%%%%%%%%%%%%%%%%%%%
\section{Conclusions and outlook}
\label{Conclusions}
%%%%%%%%%%%%%%%%%%%%%%%%%%%%%%%%%%%%%%%%%%%%%%%%%%%%%%%%%%%%%%%%%%%%%%%%%%%%%%%%%%%%%%%%%%%%%%%%%%%%%%%%%%%%%%%%%%%%%%%%%%

We discussed a conceptually simple realization of a phase space whose variables close a deformed Heisenberg algebra
\begin{equation}\label{full_alg}
[X^{i},P^{j}] =  \ii \hbar\,F^{i\,j}(\vec{P}) \;, \ \  [X^{i},X^{j}] = \ii\theta^{ij}  \;, \ \   [P^{i},P^{j}] = 0 \,.
\end{equation}
Deformations of the Heisenberg algebra are usually studied in the formalism of Hopf algebras and noncommutative geometry \cite{majidqg,connesnc}. The deformation given above has not been studied yet in this context. It would be interesting, from a mathematical point of view, to investigate the elements of the noncommutative geometry, such as co-product, star product, and the deformation of the energy-momentum conservation (the momenta addition rule). From a physical point of view, the most interesting question is that of the structure of space-time following from the full algebra (\ref{full_alg}), given that the first commutator, while keeping the other two trivial, gives rise to a minimal length \cite{AliDasVagenas1}, whereas the second commutator, while keeping the remaining two trivial, leads to the uncertainty compatible with the constraints on the position measurement that are imposed by the (classical) gravity \cite{Doplicher1995,meadgravity}.

Albeit powerful and elegant, though, the mathematical formalism alone cannot resolve the inherent ambiguities when trying to define dynamics that would correspond to the given particular deformation. Given the results of this letter, graphene could be used not only to confirm theoretical models, but to guide the construction of the physical theory itself, by providing experimental constraints. Indeed, the Dirac-like dynamics resists beyond the lowest energy approximation of the conductivity electrons of graphene, hence a ``quantum-gravity-corrected'' emergent field theory is there at our disposal. One must simply use new phase-space variables $(\vec{X},\vec{P})$ that enjoy the generalized Heisenberg algebra \eqref{full_alg}, where noncommuting coordinates, $\vec{X}$, obeying $[X^{i},X^{j}] = \ii \, \ell^2 \,  \epsilon^{i j}$, are very natural.

On the one hand, the effect here is enormous compared to the one that might take place at the fundamental level, $\Delta X^{1}\, \Delta X^{2} \ge \ell^2 / 2 \sim 10^{50} \, \, \ell_{Planck}^2$. On the other hand, there is no need for a magnetic length, $\ell_B = \sqrt{\hbar c / e B}$, as often required for noncommutativity in similar set-ups \cite{Jackiw}, nor do we need localization processes that create micro black holes \cite{Doplicher1995}.

As such, our findings are very general, macroscopic, and easy to obtain. Hence, they offer a reliable and robust way to test noncommutative theories in tabletop analog experiments. Here we propose both the general recipe for the practical implementation of this approach and a specific arrangament that illustrates how to apply that recipe.

\section*{Acknowledgements}
We are indebted to Salvatore Mignemi for the many discussions on noncommutativity and related topics. A.~I. and P.~P. gladly acknowledge support from Charles University Research Center (UNCE 24/SCI/016). P.~P. is also supported by Fondo Nacional de Desarrollo Cient\'{i}fico y Tecnol\'{o}gico--Chile (Fondecyt Grant No.~3200725).

%%%%%%%%%%%%%%%%%%%%%%%%%%%%%%%%%%%%%%%%%%%%%%% Bibliography %%%%%%%%%%%%%%%%%%%%%%%%%%%%%%%%%%%%%%%%%%%%%%%%%%%%%%%%%%%%%
\bibliographystyle{apsrev4-2}
\bibliography{three_layers_GUP_paper_biblio}
%%%%%%%%%%%%%%%%%%%%%%%%%%%%%%%%%%%%%%%%%%%%%%%%%%%%%%%%%%%%%%%%%%%%%%%%%%%%%%%%%%%%%%%%%%%%%%%%%%%%%%%%%%%%%%%%%%%%%%%%%%

\end{document}